\documentclass[11pt,draftcls,onecolumn,journal]{IEEEtran}
\usepackage{xcolor}
\definecolor{revisionpurple}{RGB}{128,0,128}
\usepackage{longtable}
\usepackage{tabularray}
\usepackage{hyperref}
\usepackage{textcomp}
\usepackage{url}
\usepackage{verbatim}
\usepackage{subcaption}
\usepackage[ruled,lined,linesnumbered]{algorithm2e}
\usepackage{amsfonts}
\usepackage{amsmath}
\usepackage{amssymb}
\usepackage{amsthm}
\usepackage{array}
\usepackage{tabularx}
\usepackage{bigints}
\usepackage{booktabs}
\usepackage{cite}
\usepackage{cleveref}
\usepackage{color}
\usepackage{diagbox}
\usepackage{epsfig,latexsym}
\usepackage{epstopdf}
\usepackage{graphicx}
\usepackage{fancyhdr}
\usepackage{float}
\usepackage{flushend}
\usepackage{indentfirst}
\usepackage{lastpage}
\usepackage{makecell}
\usepackage{mathtools}
\usepackage{multirow}
\usepackage{psfrag}
\usepackage{setspace}
\usepackage{stfloats}
\usepackage{subfloat}
\usepackage{tikz}
\usepackage{times}
\usepackage{bm}
\usepackage{breqn}
\usepackage{tablefootnote}
\usepackage{braket}

\usepackage{tcolorbox}
\makeatletter
\let\NAT@parse\undefined
\makeatother
\usepackage{hyperref}  
\def\BibTeX{{\rm B\kern-.05em{\sc i\kern-.025em b}\kern-.08em
    T\kern-.1667em\lower.7ex\hbox{E}\kern-.125emX}}
\usepackage{balance}

\usepackage{tikz}
\usetikzlibrary{arrows, calc, decorations.markings, positioning}
\usepackage{nomencl}
\usepackage{etoolbox}

\allowdisplaybreaks[2]
\begin{document}
\title{\vspace{-0em}\LARGE Technical Report\\ ``OFDM-Assisted Simultaneous Quantum and Classical THz Communications"}
\author{ Xin Liu, \textit{Graduate Student Member, IEEE}, Chao Xu, \textit{Senior Member, IEEE}, Soon Xin Ng, \textit{Senior Member, IEEE}, Mohammed El-Hajjar, \textit{Senior Member, IEEE}, Phuc V. Trinh, \textit{Senior Member, IEEE}, Shinya Sugiura, \textit{Senior Member, IEEE}, and Lajos Hanzo, \textit{Life Fellow, IEEE}\vspace{-0em}
\thanks{Xin Liu, Chao Xu, Soon Xin Ng, Mohammed El-Hajjar, and Lajos Hanzo are with
	the School of Electronics and Computer Science, University of Southampton,
	SO17 1BJ Southampton, U.K. (e-mail: xl17g20@soton.ac.uk; cx1g08@soton.ac.uk; sxn@ecs.soton.ac.uk; meh@ecs.soton.ac.uk; lh@soton.ac.uk). Phuc V. Trinh and Shinya Sugiura are with Institute of IndustrialScience, The University of Tokyo, Tokyo 153-8505, Japan. (e-mail: trinh@iis.u-tokyo.ac.jp; sugiura@iis.u-tokyo.ac.jp).
	}
}
\maketitle



\begin{abstract}
	The feasibility of cost-effective simultaneous quantum and classical communication (SQCC) transmitting both the quantum key and classical information via a superimposed coherent state is investigated both in optical and Terahertz (THz) bands.  Since the existing THz SQCC schemes assume single-carrier (SC) transmission over flat fading channels, we embark on 
investigating SQCC in realistic frequency-selective multipath THz fading channels.
We then propose an orthogonal frequency division multiplexing (OFDM) based SQCC system for time-invariant frequency-selective THz scenarios, supported by low-density parity-check coded (LDPC) multidimensional QKD reconciliation schemes.
Our simulation results demonstrate that the OFDM-based SQCC scheme is capable of achieving a practical secret key rate (SKR) over a wide range of power sharing scenarios between the classical and quantum signals. By contrast its  single-carrier counterpart requires the classical signal to be at least 100 times stronger than the quantum signal in THz SQCC.
\end{abstract}


\section{Introduction}\label{intro}
Secure quantum key distribution (QKD) is attracting significant research attention \cite{10465670}.
However, the discrete variable QKD (DV-QKD) schemes, such as the classic Bennett-Brassard-1984 (BB84) protocol \cite{BB84}, require high-cost single-photon source and single-photon detector. As a remedy, continuous variable QKD (CV-QKD) arrangements, such as the Grosshans-Grangier-2002 (GG02) protocol \cite{grosshans2002continuous} are capable of operating by relying on either homodyne or heterodyne detection, hence they are more compatible with the operational network infrastructure \cite{Weedbrook2012}.
As a parallel trend, to meet the explosive data demand of next-generation (NG) systems, the Terahertz (THz) band has been extensively studied in the communication community \cite{8663550}.
In light of this, the implementation of CV-QKD in the THz band has recently attracted considerable attention \cite{Ottaviani2020, Liu2018}. This is inspired by the fact that the THz band is more robust to the presence of dust, fog, and atmospheric turbulence conditions compared to free space optical scenarios, albeit its particle-like behaviour is significantly less pronounced than that of photonic communications.
Furthermore, multiple-input multiple-output (MIMO) techniques and orthogonal frequency division multiplexing (OFDM) have been harnessed for CV-QKD in the THz band  \cite{kundu2021mimo,Liu2021b} in order to mitigate the severe path loss and the detrimental multipath effect.
Moreover, an orthogonal time frequency space (OTFS) modulation based and low-density parity-check (LDPC) coding assisted multidimensional reconciliation (MDR) CV-QKD scheme was conceived in \cite{liu2025otfs_qkd,11069272} for transmission over doubly selective fading THz channels. These evolutionary steps might pave the way for future CV-QKD studies in the high-mobility scenarios of  NG space-air-ground integrated networks (SAGIN) \cite{9874856}.

{\color{black}{In the current era, the integration of quantum communications into NG wireless networks is becoming essential for ensuring information-theoretic security in the face of emerging quantum computing threats. 
		However, existing architectures typically employ classical communication (ClC) and quantum communication (QuC) systems separately, which requires dedicated transceivers and independent transmission resources. This duplication leads to increased hardware cost, power consumption, and system complexity, which is particularly undesirable for resource-constrained platforms. 
		To address these limitations,  simultaneous quantum and classical communication (SQCC) has recently emerged as a promising design paradigm, where both classical information and quantum states are jointly mapped onto the same physical signal and transmitted using a unified transceiver, thus offering improved spectral efficiency, reduced hardware redundancy, and enhanced compatibility with existing communication infrastructures, compared to conventional ClC and QuC coexistence schemes that rely on time \cite{xu2023simultaneous}, frequency \cite{shao2025integration}, or wavelength \cite{schreier2023coexistence,mantey2025coexistence} based separation.
		Some SQCC based protocols have already been studied~\cite{qi2016simultaneous,qi2018noise,pan2020simultaneous,tan2024simultaneous,kumar2019experimental,xu2023simultaneous,liu2023continuous,li2024passive}, which are summarized in Table~\ref{table:gap_analysis_OFDM_OTFS_SVD}.}}
More specifically, Qi firstly proposed a coherent state-based SQCC system, where both the bits destined for classical communication and the Gaussian modulated coherent states (GMCS) used for CV-QKD are mapped onto the same weak coherent pulse and decoded by the same coherent receiver~\cite{qi2016simultaneous}. It was demonstrated that both deterministic classical communication having a sufficiently low bit error rate (BER) and an adequate secret key rate (SKR) can be achieved simultaneously for transmission over tens of kilometres of single-mode fibres. Furthermore, a refined noise model of the SQCC system was established in~\cite{qi2018noise}. 
Following these developments, Pan \textit{et al.} \cite{pan2020simultaneous} proposed an advanced measurement-device-independent (MDI) SQCC protocol, which can mitigate the risk of eavesdropping associated with imperfect detectors. The simulation results demonstrated that it is feasible to operate the MDI-based SQCC system proposed over a 21~km optical fibre section by using superposition modulation. 
As a further advance, Tan \textit{et al.}~\cite{tan2024simultaneous} proposed 
a SQCC protocol based on pulse position modulation (PPM), where the ideal time slots in classical PPM are utilized for GMCS-based CV-QKD.
The so-called GMCS-PPM-SQCC avoids direct superposition of the classical and quantum signals, which reduces their mutual interference, hence leading to a longer secure distance than SQCC.
Kumar \textit{et al.}~\cite{kumar2019experimental} carried out an experimental demonstration of SQCC by achieving 25~km secure transmission via optical fibre.
Xu \textit{et al.} \cite{xu2023simultaneous} experimentally verified the feasibility of SQCC in optical fibre transmission over a shared infrastructure simultaneously, achieving a communication bit rate of 200~Mbits/s and a SKR of 1.13~Mbits/s  over a fibre having 6~dB attenuation.
Furthermore, a SQCC-CV-QKD system operating in the THz band was conceived in \cite{liu2023continuous,li2024passive}, where both atmospheric loss-based  and inter-satellite-based THz channels were considered.

{\color{black}{Nonetheless, THz-band SQCC research is still in its infancy, since numerous open problems have not yet been addressed. Firstly, although CV-QKD has been tentatively tested in wireless THz  communication, the existing SQCC studies  mainly focus on optical fibre. Recent investigations of  THz SQCC~\cite{liu2023continuous,li2024passive} only considered flat fading.
		At the time of writing, the channel dispersion of wireless THz channels has not been investigated in SQCC\footnote{\color{black}{Mobility, pointing errors, beam misalignment, and beam-squint
			may also critically affect highly directional THz transmission
			\cite{11309975,saeidi2025resourceallocationcooperativemidbandthz,11039126}.
			In this work, however, we deliberately focus on multipath-induced
			channel dispersion as a first extension of existing THz SQCC-CV-QKD
			studies, while using the atmospheric-loss- and inter-satellite-based
			scenarios considered in \cite{liu2023continuous,li2024passive} as
			literature-based benchmarks. The individual and joint effects of the other
			practical THz impairments are left for future investigation.}}.  Secondly, THz beamforming has never been harnessed for compensating the severe path loss in THz SQCC applications.
		Thirdly, the power sharing between classical and quantum components has a substantial impact on both functionalities, which has never been investigated in the context of LDPC CV-QKD reconciliation schemes. Against this background, an OFDM-based THz SQCC system operating in the face of frequency-selective fading is conceived. The novel contributions of this work are as follows:}}
\begin{itemize}
	\item A single-carrier (SC)-based SQCC system is constructed for frequency-selective fading THz channels, where the classical modulated signal used for communication and the Gaussian modulated (GM) CV-QKD signal are superimposed and transmitted in the time-domain (TD).
	\item An OFDM-based multi-carrier SQCC system is also conceived for mitigating the detrimental multipath effects of frequency-selective THz fading, where the classical modulated signal used for communication and the GM CV-QKD signal are superimposed in the frequency-domain (FD).
	\item We then intrinsically amalgamate  beamforming with both the SC and OFDM SQCC systems for mitigating the severe THz path-loss. Naturally,  the imperfect demodulation of the strong classical signal component contaminates the equivalent quantum channel (QuC), while the CV-QKD signal perturbs the classical demodulation. Both of these must be mitigated.
	\item To deal with this mutual interference, we find the most appropriate power sharing ratio between the classical and quantum signal components followed by cancelling their interference.  Explicitly, the stronger classical signal is detected first, which is only marginally contaminated by the weaker quantum signal. Hence it has high detection integrity. The detected signal is then remodulated and subtracted from the composite signal, leaving the decontaminated CV-QKD signal behind. If the classical signal imposed excessive interference, the CV-QKD reconciliation may report the presence of eavesdropping and the process must be reinitiated. 
	\item We demonstrate that the OFDM-based SQCC scheme is capable of achieving an adequate SKR over a wide range of power sharing ratios between the classical and quantum signals, whereas its SC-based counterpart requires the classical signal to be at least 100 times stronger than the quantum signal in THz SQCC.
\end{itemize}
\begin{table}[tbp]
	\scriptsize
	\centering
	\caption{Novel contributions of this work contrasted to the state-of-the-art SQCC schemes.}
	\begin{tabular}{|m{0.1\textwidth}|c|c|c|c|c|}
		\hline
		Contributions              & \textbf{This work} &\cite{qi2016simultaneous,qi2018noise,pan2020simultaneous,tan2024simultaneous}  &\cite{liu2023continuous}&\cite{kumar2019experimental,xu2023simultaneous}&\cite{li2024passive} \\\hline \hline
		Optical fibre &  &$\checkmark$& &$\checkmark$&\\\hline
		Terahertz & $\checkmark$ & &$\checkmark$ &&$\checkmark$\\\hline		
		SISO&$\checkmark$ &$\checkmark$& $\checkmark$&$\checkmark$&$\checkmark$\\\hline
		MIMO& $\checkmark$&&  &&\\\hline
		Beamforming &  $\checkmark$&&   &&\\\hline
		Atmospheric loss       &  $\checkmark$ && &&$\checkmark$\\\hline
		Inter-satellite  &   && $\checkmark$&&$\checkmark$\\\hline
		Frequency-selective fading&  $\checkmark$&& &&\\\hline
		Experimental verification           &  && &$\checkmark$&\\\hline
		OFDM           &  $\checkmark$ && &&\\\hline		
		Power sharing           &  $\checkmark$ && &&\\\hline		
	\end{tabular}
	\label{table:gap_analysis_OFDM_OTFS_SVD}
\end{table}	

{\color{black}{
\section{SC-based SQCC Protocol in Frequency Selective THz Channels}
\begin{figure}[tbp]
\centering
\begin{subfigure}{0.2\columnwidth}
\centering
\includegraphics[width=.99\linewidth]{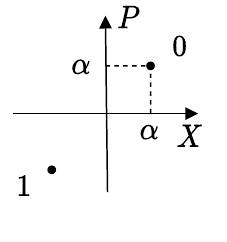}
\caption{}
\end{subfigure}
\hfill
\begin{subfigure}{0.2\columnwidth}
\centering
\includegraphics[width=.99\linewidth]{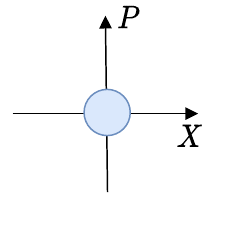}
\caption{}
\end{subfigure}
\hfill
\begin{subfigure}{0.2\columnwidth}
\centering
\includegraphics[width=.99\linewidth]{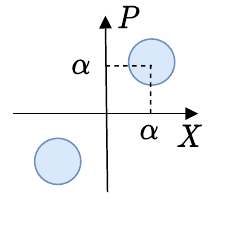}
\caption{}
\end{subfigure}	
\caption{Phase-space representations of various communication/quantum modulation schemes: (a) Classical BPSK scheme; (b) GM QKD scheme; (c) SQCC scheme.}
\label{SQCC_modulation_scheme}
\vspace{-0em}
\end{figure}
\subsection{Representations of communication/quantum modulation schemes}
As shown in Fig.~\ref{SQCC_modulation_scheme},  there are three different phase-space representations of various communication/quantum modulation schemes, namely the classical binary phase shift keying (BPSK), GM QKD and SQCC schemes, respectively.
To elaborate further, in the classical BPSK modulation scheme shown in Fig.~\ref{SQCC_modulation_scheme}(a), the classical information bits are mapped to the phase of a transmitted signal, which can be expressed as $e^{-i(m_A-1/4)\pi}\alpha$, where $m_A=0,1$ represents the information bits, and $\alpha$ is the modulation amplitude. Naturally, we could also use $\left[0,\pi\right]$ instead of $\left[\pi/4,-3/4\pi\right]$.
On the other hand, in GM CV-QKD shown in Fig.~\ref{SQCC_modulation_scheme}(b), Alice prepares coherent states $\left|x_A+ip_A\right\rangle$, which are constituted by a pair of independent Gaussian distributed random variables, denoted as  ${x}_{A}, {p}_{A} \sim \mathcal{N}\left({0}, V_{s} \right)$, where $V_s$ is the variance of the Gaussian signal.
In contrast to the classical BPSK modulation scheme and GM QKD scheme, the SQCC protocol harnesses the superposition of the classical modulation and  QKD GM, as shown in Fig.~\ref{SQCC_modulation_scheme}(c). The superimposed {\color{black} Gaussian-modulated displaced thermal} states are denoted as $\left|\left(x_A+e^{-im_A\pi}\alpha\right)+i\left(p_A+e^{-in_A\pi}\right)\alpha\right\rangle$, where $m_A=n_A\in{0,1}$ for BPSK, and $\{m_A,~n_A\}\in{0,1}$ for quadrature phase shift keying (QPSK).

\begin{figure}[tbp]
\centering
\includegraphics[width=.8\linewidth]{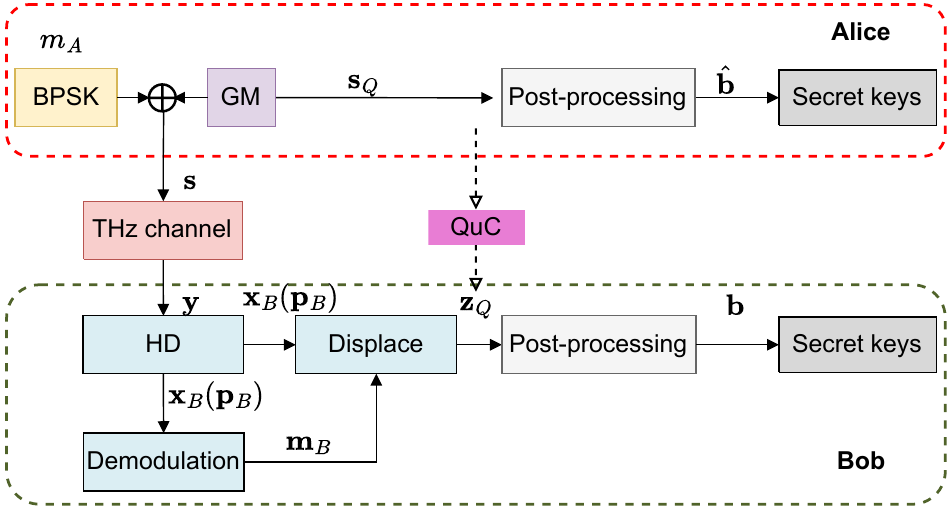}
\caption{SC based SQCC protocol diagram in THz transmission, where the reconciliation post-processing is detailed in \cite{liu2024road}. GM: Gaussian modulation; HD: homodyne detection.}
\label{fig:single_carrier_SQCC}
\vspace{0em}
\end{figure}
\subsection{SC-based SQCC scheme in frequency selective THz fading channels}
Fig.~\ref{fig:single_carrier_SQCC} illustrates the SC-based SQCC protocol in frequency selective THz channels\footnote{It is widely acknowledged that particle-like behaviour of photons is gradually eroded in the THz radio frequency (RF) band, where predominantly wave-like behaviour prevails \cite{weedbrook2012continuous}. Nonetheless, a non-zero SKR was still heralded in \cite{weedbrook2012continuous}. Hence a RF THz channel is considered in our treatise.}, which shows the superimposed classical and quantum modulated signals.
Firstly, we consider a bandwidth of $B$ and transmission frame duration of $T_f$.  {\color{black}{The feasibility of thermal-state CV-QKD at THz frequencies follows the channel assumptions established in \cite{weedbrook2012continuous,Ottaviani2020}. In SC systems, Alice prepares a Gaussian-modulated displaced thermal state $\ket{s_i}$ in the TD,
		whose complex displacement is
		$s_i=\left(x_A^i+e^{-im_A^i\pi}\alpha\right)
		+j\left(p_A^i+e^{-in_A^i\pi}\alpha\right)$,
		with $i\in[0,I-1]$, where $I$ represents the number of symbols transmitted within a frame, while symbol and frame durations are given by $T_s=1/B$ and $T_f=I\cdot T_s$, respectively.}}
For a MIMO THz scheme using $N_{Tx}$ transmit antennas (TAs) and $N_{Rx}$ receive antennas (RAs), the TD fading matrix is modelled by \cite{9874856}:
\begin{equation}
\mathbf{H}_{i,l}=\sqrt{N_{Tx}N_{Rx}}~\cdot\sum_{p=0}^{P_l-1}\widetilde{h}_p\omega_{I}^{k_p\left(i-l_p\right)}\mathbf{a}_{Rx}(\theta_{Rx,p})\mathbf{a}_{Tx}^H(\theta_{Tx,p}),
\end{equation}	
where there are $P_l$ paths falling into the $l$th time delay bin, while $\widetilde{h}_p$, $k_p$ and $l_p$ represent the fading gain, the Doppler index and delay index of the $p$th path, respectively. 
Note that a total number of $P$ paths fall into $L$ resolvable delay bins, i.e. we have $P=\sum_{l=0}^{L-1}P_l$.
Furthermore,
$\mathbf{a}_{Tx}(\theta_{Tx,p})$ and $\mathbf{a}_{Rx}(\theta_{Rx,p})$ represent the antenna responses of the TAs and RAs associated with the angle of departure (AoD), angle of arrival (AoA) $\theta_{Tx,p}$ and $\theta_{Rx,p}$, respectively.
Therefore, the analog beamformed faded channel impulse response (CIR) is expressed as \cite{liu2025otfs_qkd}:
\begin{equation}\label{ABFed_channel_ABF_single_carrier}
h_{i,l}^{RF}=\left(\mathbf{w}^{Rx,RF}\right)^H\mathbf{H}_{i,l}\mathbf{w}^{Tx,RF},
\end{equation}
where $\mathbf{w}^{Tx,RF}\in\mathcal{C}^{N_{Tx}\times 1}$ and $\mathbf{w}^{Rx,RF}\in\mathcal{C}^{N_{Rx}\times 1}$ are tuned to the LoS antenna response vectors.

Therefore, the received signal after analog combining is 
\begin{equation}				y_{i}=\sqrt{T}\sum_{l=0}^{L-1}h_{i,l}^{RF}s_{i-l}+v_{i},\end{equation} with $v_{i}=\sqrt{T}\sum_{l=0}^{L-1}h_{i,l}^{RF}{s_0}_{i-l}+\sqrt{1-T}{s_E}_{i}$, 
{\color{black}{where $T$ represents the distance-dependent transmissivity of the
lossy Gaussian quantum channel, while $s_0$ denotes the preparation
thermal-noise component and $s_E$ denotes the environmental mode
accessible to Eve under the collective Gaussian-attack model
\cite{kundu2021mimo,Ottaviani2020}.}}
Hence, the received TD signal can be expressed in vectorial form as
\begin{equation} \mathbf{y}=\sqrt{T}\mathbf{H}^{RF}\mathbf{s}+\mathbf{v},\end{equation} 
where
$\mathbf{H}^{RF}\in\mathcal{C}^{I\times I}$ represents the faded CIR\footnote{{\color{black}The classical THz propagation model is not used as a substitute for the quantum channel. Since the Gaussian-modulated quantum signal is carried by coherent electromagnetic states, its field amplitudes, or equivalently its quadrature first moments, undergo the same linear transfer response as a classical electromagnetic field in a passive linear THz medium \cite{weedbrook2012continuous,Ottaviani2020}.}}

We assume that Bob uses a homodyne detector (HD) for measuring either the $X$ quadrature or the $P$ quadrature and observes either $x_B^i=\mathfrak{Re}\left[y_i/{h}_{i,0}^{RF}\right]$ or $p_B^i=\mathfrak{Im}\left[y_i/{h}_{i,0}^{RF}\right]$, where ${h}_{i,0}^{RF}={\mathbf{H}}^{RF}[i,i]$.
Again, the stronger classical signal is demodulated first based on $x_B^i$ or $q_B^i$. Explicitly, the classical bit $m_B^i$ is deemed to be 0 if  $x_B^i(q_B^i)>0$, otherwise $m_B^i=1$ is inferred.
Following this, the measurement results are displaced according to Fig.~\ref{SQCC_modulation_scheme}(c) based on the demodulated classical signal, leading to the raw sequence of either $z_Q^i=x_B^i+\left(2m_B^i-1\right)\alpha$ or $z_Q^i=p_B^i+\left(2m_B^i-1\right)\alpha$.
The classical demodulation is assumed to be public, but the weak pulse based Gaussian sequence $\mathbf{z}_Q=[z_Q^0,z_Q^1,...,z_Q^{I-1}]^T\in\mathcal{R}^{I\times 1}$ takes advantage of the associated quantum noise to preserve uncertainty, as intimated by the equivalent QuC in Fig. 2. Naturally, the impact of imperfect demodulation will be coupled into the QuC, and similarly the GM imposes extra impairments on the classical demodulation. Finally, as seen in Fig.~\ref{fig:single_carrier_SQCC}, QKD post-processing including sifting, reconciliation and privacy amplification \cite{liu2024road,liu2025otfs_qkd,11087626} is performed based on $\mathbf{z}_Q$, in order to extract the key $\mathbf{b}$. In this treatise, we adopt syndrome-based LDPC decoding to assist QKD reconciliation, as illustrated in \cite{liu2024road}.

\section{OFDM-based SQCC Protocol in Frequency Selective THz Channels}
\begin{figure}[tbp]
\centering
\includegraphics[width=.8\linewidth]{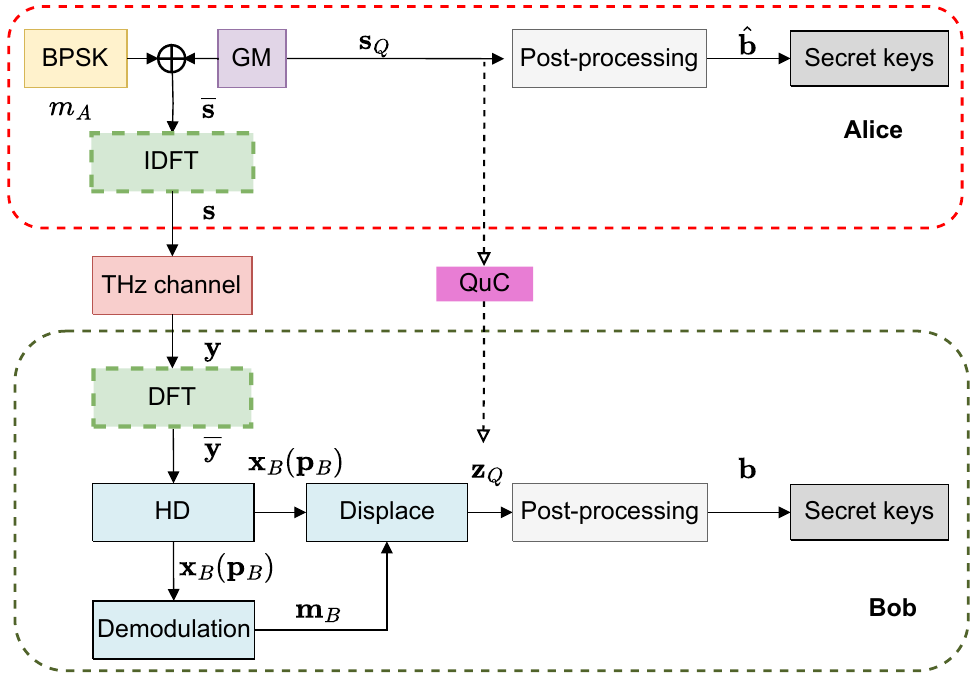}
\caption{OFDM based SQCC protocol diagram in THz transmission, where the reconciliation post-processing is detailed in \cite{liu2024road}.}
\label{fig:multi_carrier_SQCC}
\vspace{0em}
\end{figure}
In contrast to the SC based system of Fig.~\ref{fig:single_carrier_SQCC}, in this section  an OFDM-based SQCC protocol is proposed, as shown in Fig.~\ref{fig:multi_carrier_SQCC}. Explicitly, the superimposed classical and quantum signals are modulated in the FD.
To elaborate further, we consider a bandwidth of $B=M\cdot \Delta f$, where $M$ and $\Delta f$
represent the number of subcarriers and subcarrier spacing, respectively. 
Furthermore, an OFDM transmission frame has the duration $T_f=NT_{\mathrm{OFDM}}=NMT_s$, where $N$, $T_{\mathrm{OFDM}}=MT_s$, and $T_s$ represent the number of OFDM symbols in the frame, the useful OFDM symbol duration, and the sampling interval, respectively.
Alice prepares the displaced thermal state $\ket{\overline{s}_{n,m}}$, whose complex displacement is
\begin{equation}
	\overline{s}_{n,m}=\left(x_A^{n,m}+e^{-\mathrm{i}m_A^{n,m}\pi}\alpha\right)
	+\mathrm{i}\left(p_A^{n,m}+e^{-\mathrm{i}n_A^{n,m}\pi}\alpha\right),
	\label{eq:ofdm_displacement}
\end{equation}
where $n\in[0,N-1]$ and $m\in[0,M-1]$.
The OFDM transmitter then maps the data symbols $\overline{s}_{n,m}$ to the $n$th OFDM symbol and the $m$th subcarrier as
$\overline{\mathbf{s}}_n=[\overline{s}_{n,0},\overline{s}_{n,1},...,\overline{s}_{n,M-1}]^T\in\mathcal{C}^{M\times 1}$. Then the  FD signals are transformed to the TD by the inverse discrete Fourier transform (IDFT) as $\mathbf{s}_n=\mathbf{F}_M^H\overline{s}_n$, where $\mathbf{F}_{M}\in\mathcal{C}^{M\times M}$ denotes the discrete Fourier transform (DFT) matrix. 
For a MIMO OFDM THz scheme using $N_{Tx}$ TAs and $N_{Rx}$ RAs, the TD fading matrix is modelled by \cite{9874856}:
\begin{equation}
\begin{aligned}		\mathbf{H}_{n,m,l}=\sqrt{N_{Tx}N_{Rx}}~\cdot\sum_{p=0}^{P_l-1}&\widetilde{h}_p\omega_{MN}^{k_p[n(M+M_{cp})+m-l_p]}\cdot\\&\mathbf{a}_{Rx}(\theta_{Rx,p})\mathbf{a}_{Tx}^H(\theta_{Tx,p}),
\end{aligned}
\end{equation}
where 
$M_{cp}$ represents the length of the cyclic prefix (CP).
Therefore, similar to the setting in Eq.~(\ref{ABFed_channel_ABF_single_carrier}), the analog beamformed faded CIR\footnote{\color{black}{Although analog beamforming originates from conventional THz communications, it also applies to the transmission of the coherent quantum states considered here. Specifically, the quantum information is encoded in the quadrature displacement of a coherent THz electromagnetic mode. The transmit and receive beamformers perform linear phase weighting and spatial combining of this mode; hence, its quadrature first moments undergo the same beamformed transfer response \(h_{i,l}^{\mathrm{RF}}=(\mathbf w^{\mathrm{Rx,RF}})^H\mathbf H_{i,l}\mathbf w^{\mathrm{Tx,RF}}\). Beamforming therefore provides directional array and collection gain for the superimposed coherent state and improves the received quadrature SNR. It is employed as a link-enabling mechanism rather than as a new beamforming algorithm, while the transmissivity \(T\) and the environmental/Eve mode remain part of the Gaussian quantum-channel model used for security analysis.}} is expressed as \cite{liu2025otfs_qkd}:
\begin{equation}\label{ABFed_channel_ABF}
h_{n,m,l}^{RF}=\left(\mathbf{w}^{Rx,RF}\right)^H\mathbf{H}_{n,m,l}\mathbf{w}^{Tx,RF}.
\end{equation}

Therefore, the received signal after analog receiver combining is \cite{liu2025otfs_qkd}
\begin{equation}				y_{n,m}=\sqrt{T}\sum_{l=0}^{L-1}h_{n,m,l}^{RF}s_{n,<m-l>_M}+v_{n,m},\end{equation} with $v_{n,m}=\sqrt{T}\sum_{l=0}^{L-1}h_{n,m,l}^{RF}{s_0}_{n,<m-l>_M}+\sqrt{1-T}{s_E}_{n,m}$. 
Hence, the received TD signal can be expressed in vectorial form as $\mathbf{y}_n=\left[y_{n,0},y_{n,1},...,y_{n,M-1}\right]^T$, which is transformed into the FD by the DFT, as follows \cite{liu2025otfs_qkd}:
\begin{equation}	\overline{\mathbf{y}}_n=\mathbf{F}_M\mathbf{y}_n=\sqrt{T}\overline{\mathbf{H}}_{n}^{RF}\overline{\mathbf{s}}_n+\overline{\mathbf{v}}_n,\end{equation}
where the channel's frequency response (CFR) matrix is given by $\overline{\mathbf{H}}_{n}^{RF}=\mathbf{F}_M\mathbf{H}_{n}^{RF}\mathbf{F}_M^H\in\mathcal{C}^{M\times M}$,  and 
$\overline{\mathbf{v}}_n=\mathbf{F}_M\mathbf{v}_n\in\mathcal{C}^{M\times 1}$.

Following this, Bob uses HD to measure either the $X$ quadrature or the $P$ quadrature and obtains $x_B^{n,m}=\mathfrak{Re}\left[\overline{y}_{n,m}/\overline{h}_{n,m}^{RF}\right]$ or $p_B^{n,m}=\mathfrak{Im}\left[\overline{y}_{n,m}/\overline{h}_{n,m}^{RF}\right]$, which are equalized by the reciprocal of the CFR $\overline{h}_{n,m}=\overline{\mathbf{H}}_{n}^{RF}[m,m]$. We note that $\overline{\mathbf{H}}_{n}^{RF}$ is a diagonal matrix in time-invariant and frequency-selective fading scenarios.
Then Bob carries out a demodulation decision based on $x_B^{n,m}$ or $p_B^{n,m}$. Explicitly, the classical bit becomes $m_B^{n,m}=0$ if  $x_B^{n,m}(p_B^{n,m})>0$, otherwise $m_B^{n,m}=1$ is inferred.
Therefore, the GM quantum signal is obtained by removing the classical information part from $x_B^{n,m}$ or $p_B^{n,m}$, which gives  $z_Q^{n,m}=x_B^{n,m}+\left(2m_B^{n,m}-1\right)\alpha$ or $z_Q^{n,m}=p_B^{n,m}+\left(2m_B^{n,m}-1\right)\alpha$.
Therefore, Bob obtains the Gaussian sequences of $\mathbf{z}_Q=[z_Q^{0,0},z_Q^{0,1},...,z_Q^{NM-1}]^T\in\mathcal{R}^{NM\times 1}$.
Alice and Bob then carry out the classical post-processing, including reconciliation and privacy amplification \cite{liu2024road,liu2025otfs_qkd}. Furthermore, the post-equalization SIC and the reconciliation procedure both for SC- and OFDM-based SQCC are summarised in Algorithm~\ref{algo_post}.

\section{Imperfect SIC and Post-SIC Reconciliation}
{\color{blue}

\begin{algorithm}[tbp]
	\footnotesize
	\color{black}
	\DontPrintSemicolon
	\caption{{Post-equalization SIC and reconciliation procedure for SC- and OFDM-based SQCC}}
	\label{alg:post_sic_reconciliation}
	\KwIn{Index set $\mathcal K$; equalized measured-quadrature samples $\{q_{B,k}\}_{k\in\mathcal K}$; displacement amplitude $\alpha$; Alice's Gaussian reference samples $\{x_{A,k}\}_{k\in\mathcal K}$; LDPC code rate $R$.}
	\KwOut{Classical decisions $\{m_B^k\}$; post-SIC sequence $\{z_{Q,k}\}$; reconciliation BLER $P_B$; reconciliation efficiency $\beta$; diagnostic residual variance $V_{\mathrm{SIC}}$.}
	\ForEach{$k\in\mathcal K$}{
		Obtain $q_{B,k}=x_B^k$ or $q_{B,k}=p_B^k$ after SC or OFDM channel equalization\;
		\eIf{$q_{B,k}\geq 0$}{
			$m_B^k\leftarrow 0$\;
		}{
			$m_B^k\leftarrow 1$\;
		}
		Reconstruct the detected classical displacement:
		$\widehat d_{B,k}\leftarrow(-1)^{m_B^k}\alpha$\;
		Remove the reconstructed displacement:
		$z_{Q,k}\leftarrow q_{B,k}-\widehat d_{B,k}$\;
	}
	Assemble $\mathbf z_Q\leftarrow\{z_{Q,k}:k\in\mathcal K\}$ and use it as Bob's input to multidimensional reconciliation\;
	Apply LDPC decoding and record the successful and failed reconciliation blocks\;
	\label{algo_post}
\end{algorithm}
}

{\color{black}
	\subsection{Post-SIC Residual Displacement}
Let $k$ denote either an SC sample $i$ or an OFDM branch $(n,m)$ after
channel equalization, and let $m_A^k,m_B^k\in\{0,1\}$ be the transmitted
and detected classical bits on the measured quadrature. After Bob removes
the detected classical displacement, the residual component in the
reconciliation input becomes:
\begin{equation}
	r_{\mathrm{SIC},k}
	=
	\alpha\left[(-1)^{m_A^k}-(-1)^{m_B^k}\right].
	\label{eq:sic_residual}
\end{equation}
Thus, $r_{\mathrm{SIC},k}=0$ for a correct classical decision and
$r_{\mathrm{SIC},k}=\pm2\alpha$ for an erroneous decision. Defining the
per-quadrature classical bit-error probability as
$P_{\mathrm e,k}^{\mathrm C}=\Pr(m_B^k\neq m_A^k)$, the residual
mean-square value is exactly
\begin{equation}
	\mathbb E\!\left[r_{\mathrm{SIC},k}^2\right]
	=
	4\alpha^2P_{\mathrm e,k}^{\mathrm C}.
	\label{eq:sic_second_moment}
\end{equation}
For equiprobable bits and a symmetric decision channel, the residual has
zero mean; hence its variance at the same reference plane is
\begin{equation}
	V_{\mathrm{SIC},k}
	=
	4\alpha^2P_{\mathrm e,k}^{\mathrm C}
	=
	2P_C P_{\mathrm e,k}^{\mathrm C},
	\qquad P_C=2\alpha^2 .
	\label{eq:sic_variance}
\end{equation}
For QPSK, Eq.~\eqref{eq:sic_variance} uses the bit-error probability on
the measured quadrature rather than the QPSK symbol-error probability.
Because the channel coefficient has already been removed by equalization,
$V_{\mathrm{SIC},k}$ is defined directly at the post-equalization
reconciliation input and it is not divided by $T$ again.

\subsection{From Post-SIC Samples to Reconciliation and SKR}
Formulating the post-SIC record as
\begin{equation}
	z_{Q,k}
	=
	x_{A,k}
	+
	n_{\mathrm{base},k}
	+
	r_{\mathrm{SIC},k},
	\label{eq:post_sic_record}
\end{equation}
where $n_{\mathrm{base},k}$ contains the preparation-noise,
propagation-noise, and detector-noise contributions after equalization,
shows that an erroneous classical decision changes the actual sample
presented to multidimensional reconciliation and the LDPC decoder.

For system-level interpretation, the corresponding post-SIC variance and
SNR may be approximated as
\begin{equation}
	V_{\mathrm{post},k}
	\simeq
	V_{\mathrm{base},k}
	+
	V_{\mathrm{SIC},k},
	\qquad
	\mathrm{SNR}_{Q,k}^{\mathrm{post}}
	=
	\frac{V_s}{V_{\mathrm{post},k}} .
	\label{eq:post_sic_snr}
\end{equation}
This approximation is used only to interpret the system-level effect of
classical-decision errors. The residual $r_{\mathrm{SIC},k}$ is discrete
and decision-dependent, and therefore it is not treated as an independent
Gaussian excess-noise term of the physical quantum channel.

Accordingly, no SIC-specific excess-noise term is directly added to
$I_{AB}$ or $\chi_{BE}$. Instead, imperfect SIC affects the reported
finite-size SKR through the actual post-SIC sequence $\{z_{Q,k}\}$. This
sequence is used in the reconciliation process to obtain the BLER $P_B$
and the reconciliation efficiency $\beta$, and these two quantities are
then used in the finite-size SKR calculation. In this way, the influence
of the classical error probability $P_{\mathrm e}^{\mathrm C}$ is
reflected through the residual post-SIC samples and the resultant
reconciliation performance, rather than through an additional term in the
Gaussian quantum-channel model.
}
\section{Secret key rate analysis}\label{SKR}	
We apply the classic collective attack \cite{liu2025otfs_qkd} for the SKR analysis of SQCC, while taking into account the TD and FD noise sources. The SKR is expressed as:
\begin{equation}\label{eq_SKR_SVD}
	K_{\textrm{finite}}=\gamma \left(1-P_B\right)\left[\beta I_{AB}-\chi_{BE}-\triangle\left(N_{\text{privacy}}\right)\right],
\end{equation}
where $\gamma$ denotes the retained proportion of the total number of data symbols exchanged by Alice and Bob after the sifting process, while $P_B$ represents the block error rate (BLER) of the reconciliation step.
Furthermore, $I_{AB}$ is the classical mutual information between Alice and Bob based on their shared correlated data, and $\chi_{BE}$ represents the Holevo information, which represents the information Eve can extract from Bob. 
 Furthermore, $\triangle\left(N_{\text{privacy}}\right)$ denotes the finite-size correction term used to guarantee composable security under finite-length privacy amplification and chooses the same value as in \cite{liu2024road}. Additionally, the reconciliation efficiency $\beta\in[0, 1]$ is defined as\footnote{\color{black}{The reconciliation BLER $P_B$ and efficiency $\beta$ are calculated
 		from the post-SIC reconciliation sequence and its corresponding
 		operating SNR. Therefore, these quantities already reflect the
 		performance degradation caused by imperfect classical displacement
 		cancellation. The SIC residual is not introduced as a separate term in
 		the SKR expression and it is not added to the physical channel excess
 		noise used for evaluating $I_{AB}$ and $\chi_{BE}$. }}\cite{Laudenbach2018}
\begin{equation}\label{reconciliation_coefficient}
	\small
	\begin{aligned}
		\beta=\frac{R}{C}&=\frac{R}{\mathbb{E}\left[0.5\log_2\left(1+\text{SNR}^{tot}\right)\right]},
	\end{aligned}
\end{equation}
where $R$ is the coding rate, and $C$ is the one-dimensional Shannon capacity. Furthermore, $\text{SNR}^{tot}$ represents the total SNR at the receivers.

Firstly, the mutual information between Alice and Bob is given by \cite{liu2025otfs_qkd}:
\begin{equation}\label{MI_AB}
	\begin{aligned}
		I_{AB}=&\frac{1}{2} \log _2\left[\frac{\eta T (V_s+V_0)+\eta \left(1-T\right)+\left(1-\eta\right)S}{\eta T V_{\mathrm{0}}+ \eta \left(1-T\right)+\left(1-\eta\right)S}\right],		
	\end{aligned}
\end{equation} 
where $\eta$ represents the detection efficiency and $S$ stands for the variance of the trusted detector's noise \cite{Ottaviani2020}, while $T=10^{-\epsilon \mathfrak{L}/10}$ represents the distance-dependent path-loss, with $\epsilon$ defined as the attenuation and $\mathfrak{L}$ as the distance between Alice and Bob.
Furthermore, $V_s$ and $V_0$ represent the variance of the Gaussian signal and the thermal noise used in the CV-QKD modulator.  

\color{black}{Secondly, to obtain a conservative security bound for the THz thermal-state protocol, we pessimistically assume that Eve holds a purification of Alice's thermal state. Under this worst-case assumption, Eve's actual Holevo information is upper-bounded as
	$	\chi_{B E} \leq S{\left(\rho_{A B}\right)}-S{\left(\rho_{A\mid B}\right)},$
	where $S(\cdot)$ denotes the von Neumann entropy. 
	Hence, the Holevo upper bound can be calculated using the symplectic eigenvalues as \cite{Laudenbach2018}
	\begin{equation}\label{Holevo_three_eigenvalues_SVD}
		\chi_{BE} \leq G\left(\upsilon_{1}\right)+G\left(\upsilon_{2}\right)-G\left(\upsilon_{3}\right),
	\end{equation}
	where $\upsilon_1$, $\upsilon_2$ and $\upsilon_3$ are symplectic eigenvalues of $\rho_{A B}$ and $\rho_{A\mid B}$ in Eq.~(\ref{Holevo_three_eigenvalues_SVD}), and $G(*)=\left(\frac{*+1}{2}\right)\cdot\log_2\left(\frac{*+1}{2}\right)-\left(\frac{*-1}{2}\right)\cdot\allowbreak \log_2\left(\frac{*-1}{2}\right)$. After substituting Eq.~(\ref{MI_AB}) and Eq.~(\ref{Holevo_three_eigenvalues_SVD}) into Eq.~(\ref{eq_SKR_SVD}), the corresponding lower bound on the SKR can be obtained.
}}}

%
\begin{figure}[tbp]
\centering
\begin{subfigure}{0.3\columnwidth}
\centering
\includegraphics[width=1.0\linewidth]{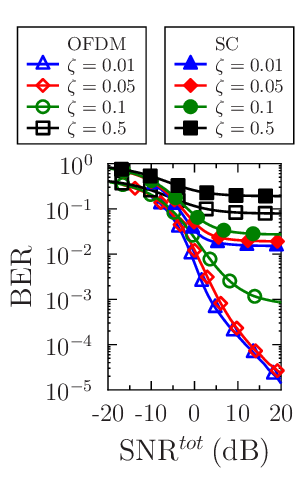}
\caption{}
\end{subfigure}
\hspace{2em}
\begin{subfigure}{0.3\columnwidth}
\centering
\includegraphics[width=1.0\linewidth]{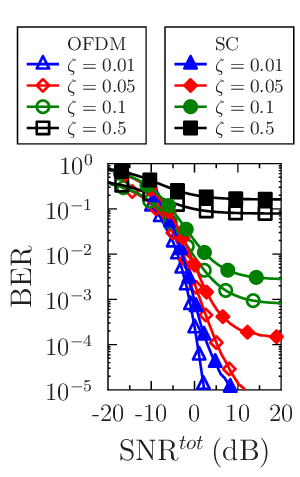}
\caption{}
\end{subfigure}
\caption{\small BER comparisons for classical communications of OFDM and SC based SQCC systems using different power sharing ratios with $4\times4$ MIMO dimensions, where (a) $K=0$~dB and (b) $K=10$~dB.}
\label{fig:SQCC_SNR_BER_classical_QPSK}
\end{figure}
\section{Performance Analysis}\label{simulation_ABF_CE}
In this section, the BLER performance of the OFDM and SC based SQCC scheme is characterized, followed by their SKR versus distance performance.
The simulation parameters are adopted from \cite{liu2025otfs_qkd} and are specified as follows.
The power of the classical QPSK modulation is $P_C=2\alpha^2$, while that of the GM quantum signal is $P_Q=2V_s$.
{\color{black} Therefore, the ratio between $P_C$ and $P_Q$ is set to $\zeta=P_Q/P_C=0.01,0.05,0.1,0.5$} based on the investigations in \cite{qi2016simultaneous,pan2020simultaneous}.
Regarding the quantum transmission channel,  the carrier frequency is set to $f_c=15$~THz, the maximum delay is set to $\tau_{max}=20$~ns, and the Ricean factor is $K=0, 10$~dB. The atmospheric absorption loss is $\epsilon=50$~dB/km, and the maximum time delay bin is $L=\lceil \tau_{max}M \Delta f\rceil=3$,  while the total number of paths is $P=L$.
The OFDM subcarrier spacing is $\Delta f=2$~MHz.
The number of subcarriers $M$ and of symbols $N$ are set to 64 and 16, respectively, while the CP length is given by $M_{cp}=L+1$.
For the SC configuration, the bandwidth is the same as in OFDM, and so is the number of modulated symbols as in OFDM, namely $I=MN$.
The number of transmitter and receiver antennas is $N_{Tx}=N_{Rx}=4$.
Furthermore, an LDPC forward error correction (FEC) code having the length of $N_{\text{FEC}}=1024$ and a coding rate of $R=0.5$ is employed for the reconciliation scheme.
{\color{black}{Additionally, a stationary scenario having the speed of $v=0$~mph is considered and the beam alignment is assumed perfect.}}
Overall, all the key parameters are summarised in Table~\ref{tab:simulation_parameters}.

{\color{black}\begin{table}[H]
	\color{black}
	\centering
	\footnotesize
	\caption{Parameters of the stationary indoor baseline.}
	\label{tab:simulation_parameters}
	\begin{tabularx}{\linewidth}{|p{0.31\linewidth}|X|}
		\hline
		\textbf{Parameter} & \textbf{Value or convention} \\
		\hline
		Carrier and propagation & $f_c=15$~THz, $\tau_{\max}=20$~ns, Ricean factor $K\in\{0,10\}$~dB, $\epsilon=50$~dB/km, and $v=0$ \\
		\hline
		OFDM waveform & $M=64$, $N=16$, $\Delta f=2$~MHz, $B=M\Delta f=128$~MHz, and $M_{cp}=L+1$ with modeled delay-bin parameter $L=3$ \\
		\hline
		SC reference & The same bandwidth and $I=MN$ transmitted symbols as the OFDM configuration \\
		\hline
		Arrays and beamforming & $N_{Tx}=N_{Rx}=4$ with fixed LoS-directed analog transmit and receive weights \\
		\hline
		Reconciliation & LDPC code length $N_{\mathrm{FEC}}=1024$ and nominal coding rate $R=0.5$ \\
		\hline
		Power-sharing sweep & $\zeta\in\{0.01,0.05,0.1,0.5\}$ \\
		\hline
		$\Delta\left(N_{\text {privacy }}\right) \approx 7 \sqrt{\frac{\log _2(2 / \epsilon)}{N_{\text {privacy }}}}$& $\epsilon$= $10^{-10}$ and $N_{\text{privacy}}$ =$10^{12}$ \\
		\hline
	\end{tabularx}
\end{table}
}

\begin{figure}[tbp]
\centering
\begin{subfigure}{0.3\columnwidth}
\centering
\includegraphics[width=1.0\linewidth]{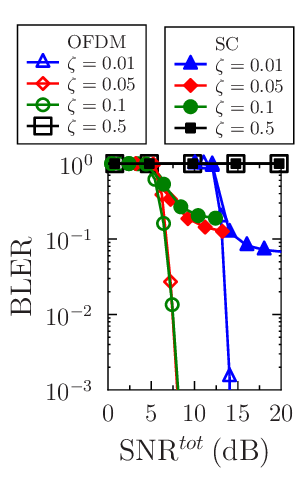}
\caption{}
\end{subfigure}
\hspace{2em}
\begin{subfigure}{0.3\columnwidth}
\centering
\includegraphics[width=1.0\linewidth]{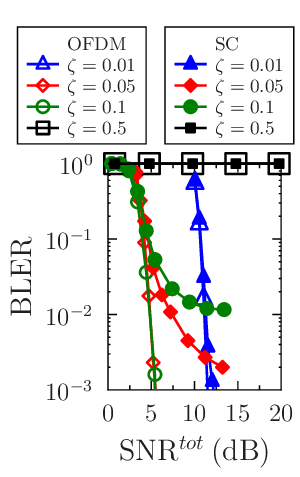}
\caption{}
\end{subfigure}
\caption{\small BLER comparisons for quantum communications of OFDM and SC based SQCC systems using different power sharing ratios with $4\times4$ MIMO dimensions, where (a) $K=0$~dB and (b) $K=10$~dB.}
\label{fig:SQCC_SNR_BER_quantum_QPSK}

\vspace{0em}
\end{figure}

{\color{black}Fig.~\ref{fig:SQCC_SNR_BER_classical_QPSK} portrays the uncoded BER performance of classical communications relying on both the SC and OFDM-based SQCC systems using $4\times4$ MIMO in the face of time-invariant frequency-selective THz fading scenarios, where the effect of power sharing ratio between the quantum and classical components is investigated.
	As seen from Fig.~\ref{fig:SQCC_SNR_BER_classical_QPSK}, the BER performance of both the SC- and OFDM- based SQCC systems improves as $\zeta$ decreases. This is because the lower $\zeta$, the higher proportion of the same total signal power is assigned to the classical signal.
	Furthermore, both the SC and OFDM based SQCC classical system performances improve as the Ricean $K$ increases, as a benefit of the improved channel quality. Nonetheless, the BER of OFDM-based SQCC systems is better than that of the SC-based SQCC systems for each of the $\zeta$ setting.
	More specifically, as demonstrated in Fig.~\ref{fig:SQCC_SNR_BER_classical_QPSK}(b), the BER drops to $10^{-5}$ at $10$~dB for the OFDM-based SQCC system having $\zeta=0.01$, whilst the BER for SC-based SQCC system has an error-floor for $\zeta=0.05$.}

Fig.~\ref{fig:SQCC_SNR_BER_quantum_QPSK} portrays the BLER of quantum communications for both the SC and OFDM-based SQCC systems using $4\times4$ MIMO in frequency-selective THz fading scenarios, where LDPC coding  is applied for reconciliation, but no FEC is used for the classical system.
As seen from Fig.~\ref{fig:SQCC_SNR_BER_quantum_QPSK}, the BLER performance of both systems improves as $\zeta$ is reduced from 0.5, 0.1 to 0.05. This is because when $\zeta$ is higher, the classical signal power becomes insufficiently high for low-BER classical demodulation, which leads to erroneous displacement for the quantum measurement and hence an error floor is observed. However, at $\zeta=0.01$, the BLER is degraded compared to $\zeta=0.05$, since the SNR experienced in the quantum domain becomes too low.
Furthermore, it can observed from Fig.~\ref{fig:SQCC_SNR_BER_quantum_QPSK}(a) that the BLER  of the OFDM-based SQCC system having $\zeta=0.1$ is the best, followed by $\zeta=0.05,0.01,0.5$. As for the SC-based SQCC system, only $\zeta=0.01$ can provide an acceptable BLER level (BLER$\leq0.1$) for SKR generation. 
However, as demonstrated in  Fig.~\ref{fig:SQCC_SNR_BER_quantum_QPSK}(b), when a higher Ricean factor of $K=10$~dB is considered, the BLER performance of both the OFDM-based and of the SC-based systems improves substantially compared to Fig.~\ref{fig:SQCC_SNR_BER_quantum_QPSK}(a). This  especially so for the SC-based scheme, hence $\zeta=0.01, 0.05, 0.1$ can provide an acceptable BLER for SKR generation. 

\begin{figure}[tbp]
\centering
\begin{subfigure}{0.5\columnwidth}
\centering
\includegraphics[width=1.0\linewidth]{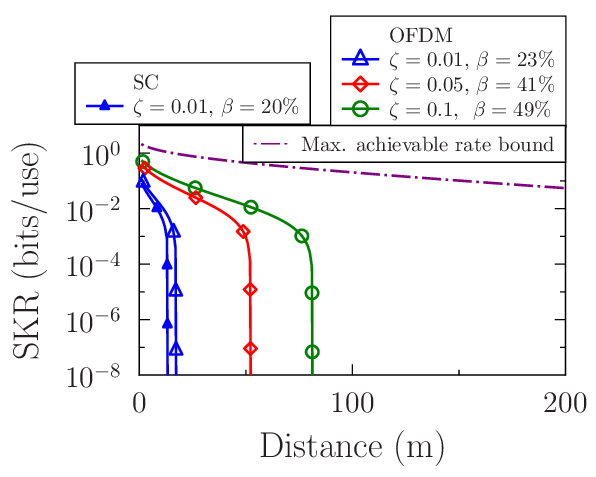}
\caption{}
\end{subfigure}
\hspace{-1em}	
\begin{subfigure}{0.5\columnwidth}
\centering
\includegraphics[width=1.0\linewidth]{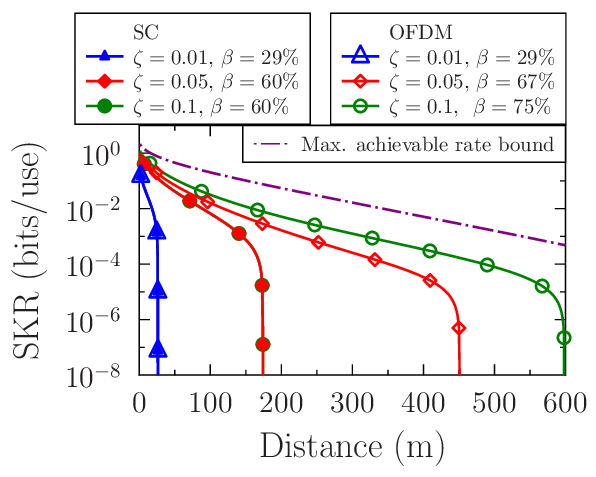}
\caption{}
\end{subfigure}
\caption{\small Lower-bound SKR performance comparison of OFDM and SC based SQCC systems using different power sharing ratios with $4\times4$ MIMO dimensions, where (a) $K=0$~dB and (b) $K=10$~dB. Note: Max. achievable rate bound corresponds to $\beta=100\%$.}
\label{fig:SKR}
\vspace{0em}
\end{figure}

Fig.~\ref{fig:SKR}(a) compares the SKR versus distance performance of SC-based and OFDM-based SQCC systems. As seen from Fig.~\ref{fig:SKR}, upon increasing $\zeta$ from 0.01, 0.05 to 0.1 in the OFDM-based SQCC system, the reconciliation efficiency $\beta$ gradually increases from $23\%, 41\%$ to $49\%$.
As a benefit, the maximum secure distance is increased.
Moreover, the SC-based system is only capable of achieving a secure distance of 12 m at $\zeta=0.01$.
Furthermore, as demonstrated in Fig.~\ref{fig:SKR}(b), the SKR versus distance performance of both the SC-based and OFDM-based SQCC system with $K=10$~dB improves compared to their $K=0$~dB counterpart characterized in Fig.~\ref{fig:SKR}(a), especially for the SC-based one, where $\zeta=0.05, 0.1$ also succeed in providing an adequate SKR.  

\section{Conclusions}

An OFDM-based SQCC system was conceived for time-invariant
frequency-selective THz fading scenarios.
{\color{black}{
			The principal insight of this work is that frequency-selective THz
propagation affects SQCC performance through both the physical channel
response and an indirect classical-decision-dependent reconciliation
pathway. The OFDM-based receiver resolves the frequency-selective
response into parallel subcarrier responses, thereby improving
classical detection and the resulting post-SIC reconciliation
performance. The calculated reconciliation BLER $P_B$ and efficiency
$\beta$ capture this improvement and are subsequently used in the
finite-size SKR evaluation. Within the evaluated stationary indoor and
aligned-beam scenario, the OFDM-based system supports reliable
reconciliation and positive SKR over a wider range of representative
classical-to-quantum power-sharing ratios than its SC counterpart. This
distinguishes the present contribution from quantum-only THz CV-QKD and
previous flat- or large-scale-loss-based THz SQCC studies.
}}
More specifically, our simulation results demonstrated that the OFDM-based SQCC system achieves a better BLER performance in  frequency-selective scenarios across a wider range of $\zeta$ compared to its SC-based counterpart. As a benefit, the reconciliation efficiency is increased, thus providing a longer secure transmission distance. {\color{black}{As a natural extension of this work, several directions merit further investigation. 
		Firstly, for time-varying and doubly selective THz channels, OTFS-based SQCC systems are of particular interest due to their inherent robustness against Doppler effects.
		Secondly, wideband THz transmission may introduce beam-squint and beam-split effects\cite{11309975,saeidi2025resourceallocationcooperativemidbandthz,11039126}, which can significantly impact beamforming performance and thus affect the overall SQCC system's behaviour. 
		Finally, practical receiver impairments, including the phase noise of the homodyne detector and the finite resolution and dynamic range of the analog-to-digital converter (ADC) should also be carefully examined, as they may introduce additional effective noise and hence influence the system design, such as the optimal power-sharing ratio.}}

\vspace{0em}
\bibliographystyle{IEEEtran}
\bibliography{Myreference_paper2_v2}
\end{document}